\documentclass[reprint,aps,prl,amsmath,amssymb,longbibliography,floatfix]{revtex4-1}

\usepackage{graphicx}
\usepackage[utf8]{inputenc}
\usepackage[T1]{fontenc}
\usepackage{dcolumn}
\usepackage{bm}
\usepackage{color}
\usepackage{braket}
\usepackage{upgreek}

\newcommand{\ia}{\ensuremath{I_\text{A}}}
\newcommand{\ib}{\ensuremath{I_\text{B}}}
\newcommand{\iab}{\ensuremath{I_\text{A/B}}}
\newcommand{\imax}{I_\text{max}}
\newcommand{\imin}{I_\text{min}}
\newcommand{\ea}{E_\text{A}}
\newcommand{\eb}{E_\text{B}}

\newcommand{\eps}{\epsilon}

\newcommand{\ada}{\ensuremath{a^{\dagger}a}}
\newcommand{\bdb}{\ensuremath{b^{\dagger}b}}
\newcommand{\adb}{\ensuremath{a^\dagger b}}
\newcommand{\eada}{\ensuremath{\exv{\ada}}}
\newcommand{\ebdb}{\ensuremath{\exv{\bdb}}}

\newcommand{\Eadb}{\ensuremath{\Exv{\adb}}}
\newcommand{\phia}{\phi_\text{A}}
\newcommand{\phib}{\phi_\text{B}}
\newcommand{\amax}{A_\text{max}}
\newcommand{\exv}[1]{\braket{#1}}
\newcommand{\Exv}[1]{\Braket{#1}}
\newcommand{\go}{\ensuremath{g^{(1)}}}

\newcommand{\absgo}{{| \go |}}

\newcommand{\iu}{\text{i}}

\renewcommand{\Re}{\operatorname{Re}}
\newcommand{\abs}[1]{\left \vert #1 \right \vert}

\allowdisplaybreaks[3]

\begin{document}

\title{Coherence and visibility in Fano interferometers}

\author{Fabian Lauble}
\author{Jörg Evers}
\affiliation{Max-Planck-Institut für Kernphysik, Saupfercheckweg 1, D-69117 Heidelberg, Germany}

\date{08.03.2017}

\begin{abstract}
Interferometry is an indispensable tool across all the natural sciences. Recently, a new type of interferometer based on phase-sensitive Fano resonances has been proposed and implemented. In these Fano interferometers, the two arms are formed by a spectrally broad continuum channel, and a spectrally narrow resonant bound state scattering channel, respectively. We show that the textbook relation between interference visibility and coherence known from double-slit- or Mach-Zehnder-interferometers does not apply to Fano interferometers, because the physical origin  of the interference extrema is different. We then show how instead the asymmetry of Fano spectra can be exploited to quantify coherence in Fano interferometers.
\end{abstract}


\maketitle

\begin{figure*}[t]
 \centering
 \includegraphics[width=2\columnwidth]{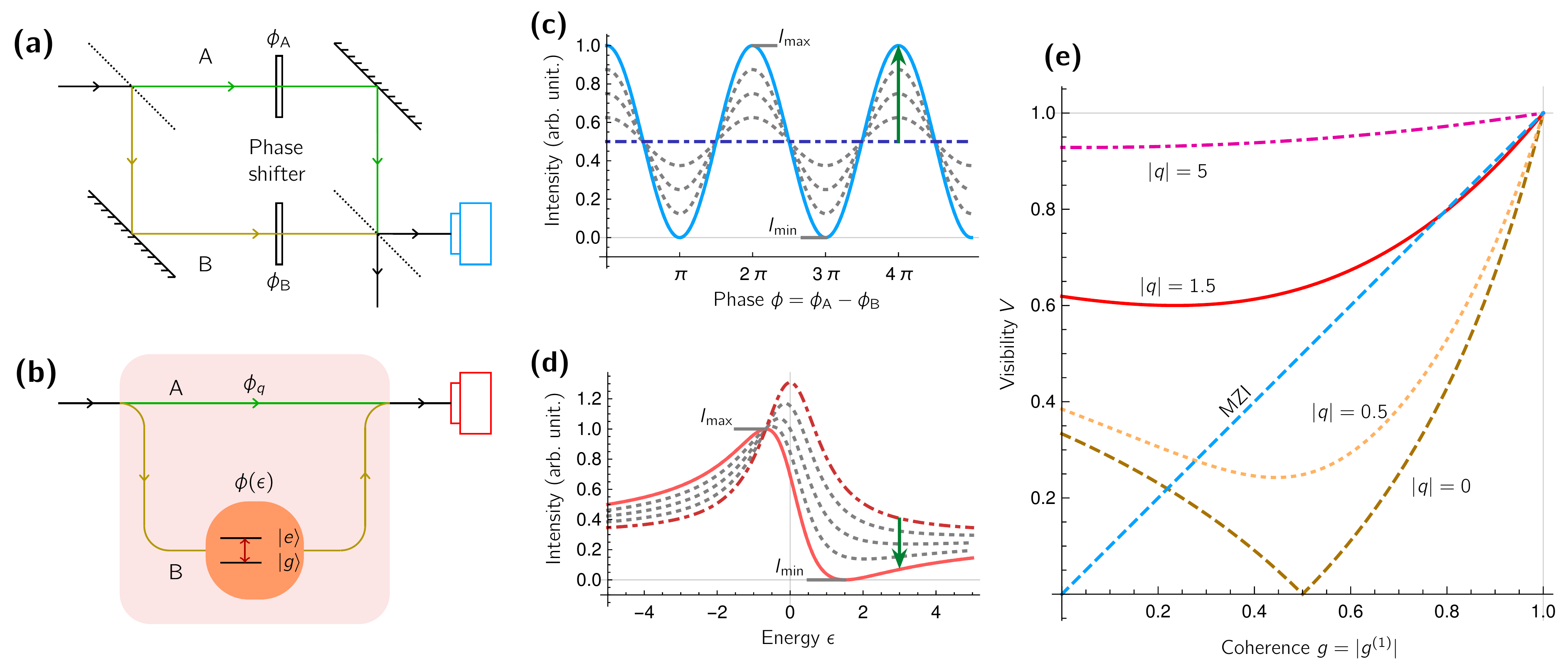}
 \caption{(Color online) %
 (a) Schematic of a Mach-Zehnder interferometer (MZI) with two paths A and B and variable phases $\phia$ and $\phib$.
 (b) Schematic of a Fano interferometer (FI) with a continuum channel with phase $\phi_q$ and a resonant channel with energy-dependent phase $\phi(\eps)$.
 (c) Interference signal $I(\phi)$ of a MZI for different values of the coherence $g$. Along the green arrow: $g=0, 0.25, 0.5, 0.75, 1$. The maxima and minima approach each other with decreasing $g$.
 (d) Fano interference spectra $I(\eps)$ for $q=-1.5$, and for the same values of $g$ as in (c). For decreasing $g$, the maximum and minimum remain different, but the asymmetric line shape transforms into a symmetric one.
 (e) Visibilities evaluated from Eq.~\eqref{eq-visibility} as functions of $g$. The straight, dashed, blue line shows the case of a balanced MZI. The other curves depict the visibility for FI for different Fano parameters $q$.
 }
 \label{fig:traditional}
\end{figure*}

Interference is a key concept in the natural sciences, which, for example, has played a major role in discovering the wave theory of light, quantum theory, special relativity and even general relativity's gravitational waves, but also is utilized by many applications~\cite{opt-int,grav,astro}.
Archetype instruments exploiting interference are double-slit (DSI) or Mach-Zehnder (MZI) interferometers and related setups~\cite{hecht,scully_quantum_1997}, in which an incoming wave is split into two channels, which after a propagation are eventually superposed again to interfere [see Fig.~\ref{fig:traditional}(a)].  It is a classic textbook calculation that the visibility
\begin{equation}
\label{eq-visibility}
V = (\imax - \imin )/(\imax + \imin )
\end{equation}
of the interference in a DSI- or MZI-like interferometer is proportional to the first-order correlation function characterizing the waves' coherence~\cite{scully_quantum_1997,walls_quantum_1994,agarwal,shih}. This on the one hand shows that coherence is a basic requirement for the occurrence of interference. 
On the other hand, interferometry is a convenient tool to measure coherence properties. This is not only useful to characterize sources of light, but also to explore the non-classicality of  light~\cite{scully_quantum_1997,agarwal}.

Recently, a different type of interferometer based on Fano interference has been put forward and experimentally demonstrated. This type of interference is usually associated with the modification of a symmetric Breit-Wigner resonance to an asymmetric Fano line shape. It appears when a resonant bound state channel interferes with a continuum channel, as schematically depicted in Fig.~\ref{fig:traditional}(b)~\cite{fano_effects_1961,limonov_fano_2017,RevModPhys.82.2257}. In Fano interferometers (FI), this interference is externally controlled and thereby exploited as a tool for interferometry. FI and related control have been discussed, e.g., in the context of quantum dots~\cite{johnson_coulomb-modified_2004}, photonic crystal circuits~\cite{miroshnichenko_machzehnderfano_2009} and transient attosecond spectroscopy~\cite{ott_lorentz_2013}. Correspondingly, decoherence can modify Fano resonances in a characteristic way~\cite{PhysRevLett.105.056801}.
FI have also been implemented in two different settings at hard x-ray photon energies, en route towards the development of x-ray quantum optics~\cite{adams_x-ray_2013}. Using planar cavities containing narrow nuclear resonances coupled near-resonantly to one of its modes, the relative phase in a quantum superposition of two nuclear states was interferometrically determined~\cite{heeg_interferometric_2015}. Second, in coherent forward scattering of broadband x-ray pulses off of large ensembles of nuclei featuring a narrow resonance, the Fano interference control was used to increase the intensity of given x-ray pulses in a narrow spectral region~\cite{heeg_spectral_2017}. 
Even though a variety of interferometers have already been demonstrated at x-ray energies~\cite{shvydko_x-ray_nodate}, these x-ray FI are of particular interest, since they neither require dedicated x-ray optics, nor elaborate alignment. Also, in FI the two interferometer arms are not spatially separated, which stabilizes the relative phase against fluctuations in their length. 

These promising developments invite an exploration of the potential of FI, in particular also related to possible test of nonclassicality. This raises the question of how the interference pattern and the coherence properties of the interfering waves are related in a FI.

Here, we show that the standard visibility known from DSI or MZI is not capable of characterizing the waves' coherence in FI because the physical origin of interference maxima is different. As our main result, we then derive an alternative method to extract and quantify the coherence properties from interference patterns measured in FI. It is  based on a parameter quantifying the asymmetry of the pattern, which is readily accessible in experiments.

{\it Coherence and visibility. }
We start by considering a generic interferometer comprising two channels A and B, with field amplitudes characterized by operators $a$ and $b$. The two channels' individual intensities are $\ia = \eada$ and $\ib = \ebdb$, respectively. The interference signal $I = \Exv{(a+b)^\dagger (a+b)}$ obtained after superposing the two channels evaluates to~\cite{scully_quantum_1997,agarwal}
\begin{align}
\label{eq-intfQuantum}
I =& \ia + \ib + 2 \Re \Eadb \\
  =& \ia + \ib + 2 \absgo \sqrt{\ia \ib} \cos \phi\,.
\end{align}
Here, the coherence of first order is
\begin{equation}
\label{eq-gone}
\go = \dfrac{\Eadb}{\sqrt{\eada \ebdb}}\, ,
\end{equation}
and the total relative phase between the two channels is $\phi=\arg \Eadb$. The coherence parameter $g = \absgo $ determines the relative strength of the interference term as compared to the individual channel intensities, and varies between full coherence ($g=1$) and no coherence ($g=0$). The case of total destructive interference, i.e. $I=0$, can only occur if $g=1$ and $\ia = \ib$.

Because the coherence $g$ controls the degree of interference, it can be determined from the interference pattern. For the case of a  MZI shown in Fig.~\ref{fig:traditional}(a), the intensity as function of the relative phase $\phi$ exhibits  interference patterns as shown in Fig.~\ref{fig:traditional}(c) for different values of $g$ from 0 to 1. The maxima and minima are due to constructive and destructive interference, characterized by $\phi = 0$ and $\phi = \pi$ (modulo $2\pi$), respectively, and can be used to quantify the interference via the visibility Eq.~(\ref{eq-visibility}). It is well known that the coherence and the visibility are related as~\citep{walls_quantum_1994,scully_quantum_1997,agarwal}
\begin{equation}
\label{vis-g}
V = \chi \, g   \, ,
\end{equation}
with $\chi = 2 \sqrt{\ia \ib} / (\ia + \ib) \leq 1$ quantifying the ratio of the two intensities $\iab$. The visibility is therefore a lower bound for the coherence and $V=g$ if and only if $\ia = \ib$. This `balanced' case is illustrated in Fig.~\ref{fig:traditional}(e). 
%

{\it Fano interferometer. }
It turns out that  Eq.~(\ref{vis-g}) only holds if the intensities $\ia$ and $\ib$ are independent of the relative phase $\phi$, as it is usually the case for MZI over a wide range of relative phases. However, in FI, one of the interfering channels is a continuum channel, while the other channel is formed by a spectrally narrow bound state resonance, see Fig.~\ref{fig:traditional}(b). Therefore, the individual channel intensities vary along with the relative phase, which gives rise to a different dependence of the interference pattern on the coherence parameter.
In the following, for definiteness of the analytical calculations, we specialize the analysis to FI based on planar cavities containing a bound state which is near-resonant to the driven cavity mode~\cite{heeg_interferometric_2015,rohlsberger_coherent_2005}. Then, the continuum channel $A$ is formed by all photons which pass through the spectrally broad cavity mode, but do not interact with the bound state resonance. All remaining possible photon pathways involving at least one interaction with the bound state form the second channel $B$. For typical cavity parameters, the amplitudes can be written as~\cite{heeg_interferometric_2015}
\begin{align}
\label{fano-amplitudes}
 \ea = \dfrac{1}{q-\iu}\,, \qquad \eb = \dfrac{1}{\eps+\iu}\,.
\end{align}
Here, the dimensionless energy $\eps = (E-E_0) / (2 \varGamma)$ characterizes the detuning of the incident light to the bound state resonance, normalized to its linewidth $\varGamma$. The Fano parameter $q$ can be controlled via the incidence angle of the light onto the cavity~\cite{heeg_x-ray_2013,heeg_interferometric_2015}. Introducing a finite coherence parameter $g$ as for the MZI, the intensity evaluates to
\begin{align}
I(\eps) &= I_A + I_B + 2\,g\,\textrm{Re}(E_A^*\,E_B) \nonumber \\
&= \dfrac{\eps^2 + q^2 + 2 q \eps g +2(1-g)}{(1+\eps^2)(1+q^2)}\, . \label{eq-FanoFiniteCoh}
\end{align}
From Eq.~(\ref{fano-amplitudes}) it can be seen that the relative phase $\phi$ between the two channels' amplitudes can be tuned via $\epsilon$, exploiting the bound state resonance's dispersion. However, in contrast to the MZI, varying $\phi$  via $\epsilon$ also changes $I_B$.

For the special case of $g=1$, we recover the standard Fano formula~\cite{connerade_interacting_1988,limonov_fano_2017}\,
\begin{equation}
\label{eq-fanoStandard}
I_{g=1} = \dfrac{(\eps + q)^2}{(1+\eps^2)(1+q^2)}\, .
\end{equation}
Example Fano line shapes are shown in Fig.~\ref{fig:traditional}(d), as function of the coherence parameter $g$. Note that the Fano spectrum is sometimes defined without the factor $(1+q^2)^{-1}$. This does not affect our results since this overall factor cancels in both the visibility Eq.~\eqref{eq-visibility} and the asymmetry parameter defined in Eq.~\eqref{eq-asymmetry} below, and has no effect on the relative phase of the amplitudes.

{\it Results. }
We start by showing  that even in the case of maximum coherence $g=1$, the nature of the interference in the FI differs from that in MZI. For this, we determine the maxima and minima of the Fano intensity Eq.~(\ref{eq-fanoStandard}), which occur at $\eps = -q$ and $\eps = 1/q$, respectively. At $\eps=-q$, the relative phase of the two channels is $\pi$, and the intensity vanishes completely, like in the MZI case. But the maximum $\eps = 1/q$ corresponds to a relative phase unequal to zero, in contrast to MZI. It is therefore not a result of perfect constructive interference, but rather originates from a different mechanism, namely a compromise between partial constructive interference and the $\eps$-dependent channel intensities. 

Next, we show that the standard visibility Eq.~\eqref{eq-visibility} cannot be used to infer the coherence from the interference signal. This is illustrated in Fig.~\ref{fig:traditional}(e), which depicts the visibility as function of the coherence parameter $g$ for different Fano parameters $q$. 
It can be seen that in the absence of coherence $g=0$, the visibility is medium or even high, depending on $q$. The reason for this is that the incoherent addition of the two channels' intensities is not constant as function of $\eps$, due to the narrow bound state channel resonance, such that $\imax \neq  \imin$ is possible independent of $g$. 
Also, the visibility completely vanishes for $q=0$ and $g=1/2$. In this case, the intensity does not depend on the relative phase of the two channels, despite the presence of partial coherence. This unintuitive feature arises since for $q=0$ and $g=1/2$ the energy dependence of the interference term exactly cancels the energy dependence of the bound state channel intensity. This yields a completely flat spectrum even though a resonance is involved.
Finally, we analyze the case of finite coherence $g$, and find that the visibility cannot uniquely be converted into the value of $g$, due to the presence of minima of the visibility as function of $g$. The visibility neither provides a practical lower bound for the coherence.
We therefore conclude that the standard visibility is unsuitable to characterize FI.

\begin{figure}[t]
 \centering
 \includegraphics[width=\columnwidth]{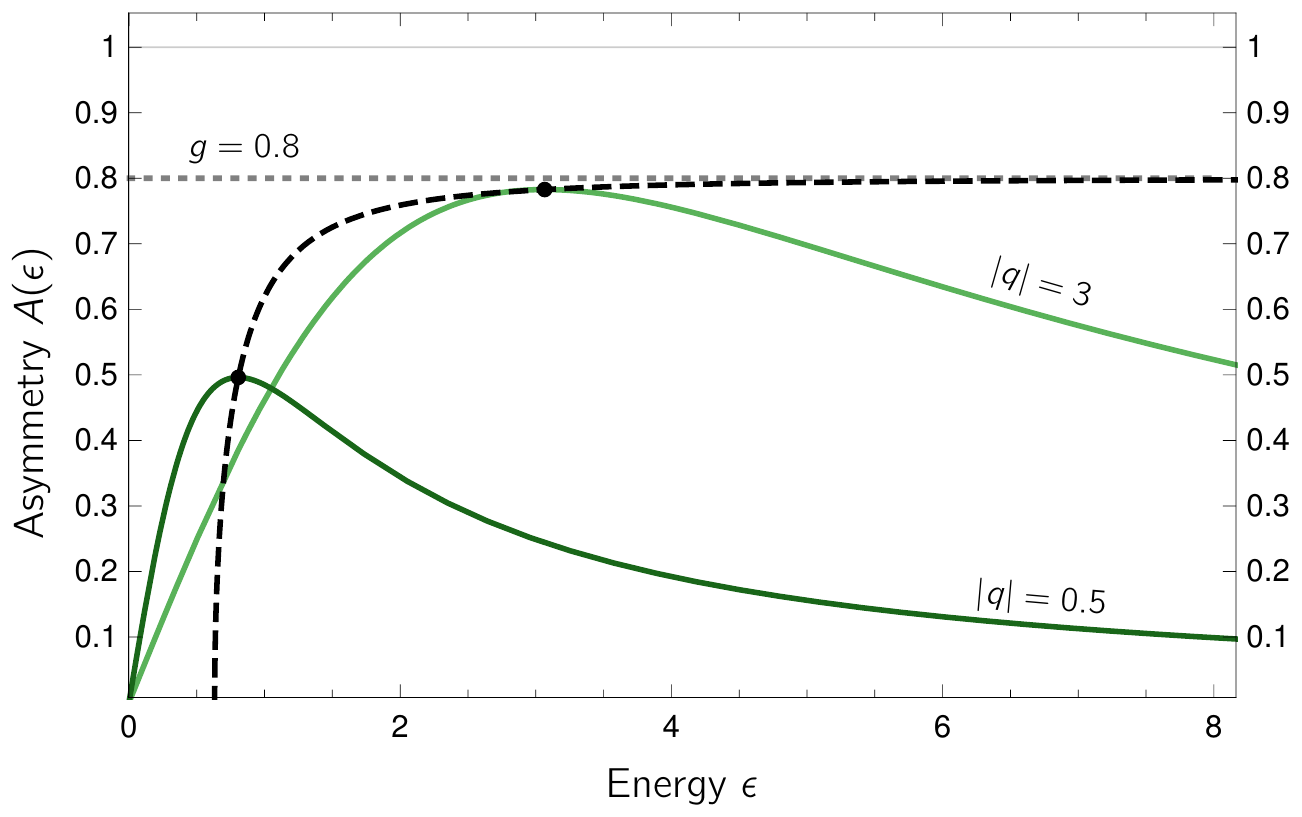}
 \caption{(Color online) Asymmetry parameter  Eq.~\eqref{eq-asymmetry-fano} for the Fano interferometer. The coherence parameter is chosen as $g=0.8$, and the two solid lines show $|q|=0.5$ and $|q|=3$, respectively. The maxima $\amax$ form lower bounds to $g$, and are marked with dots. The dashed line shows the positions $(\eps_0, \amax)$ of the maxima for other $q$ values. With increasing $|q|$, the maximum approaches $g$ as desired.
 }
 \label{fig:asymmetry}
\end{figure}

Motivated by this, we now put forward an alternative method to characterize and measure the coherence parameter in the case of FI. Our approach is motivated by the connection of the degree of interference to the coherence parameter. Analyzing $I(\eps)$, we find that the individual channels' intensities are even functions of $\eps$, while the interference part contains contributions which change their sign under the transformation $\eps \to -\eps$. This leads us to define the asymmetry parameter
\begin{align}
A(\eps) = \left\vert 
		  \dfrac{I(\eps)-I(-\eps)}{I(\eps)+I(-\eps)}
		  \right\vert \, , \label{eq-asymmetry}
\end{align}
which can be understood as the (modulus of the) ratio of the anti-symmetric to the symmetric part of $I(\eps)$. As the intensity is non-negative, $0 \leq A(\eps) \leq 1$.  Since $A(\eps)$ is symmetric in $\eps$, we will only consider $\eps \geq 0$ in the following. Inserting the Fano intensity Eq.~\eqref{eq-FanoFiniteCoh}, the asymmetry parameter evaluates to 
\begin{equation}
\label{eq-asymmetry-fano}
A(\eps) = \dfrac{2\, \lvert q \rvert \eps  }{2(1-g) + \eps^2 + q^2} \; g  \, .
\end{equation}
A closer analysis reveals that the fraction in Eq.~(\ref{eq-asymmetry-fano}) is bounded by one. As a result, we can derive a lower bound to the coherence $g$ directly from the asymmetry parameter via 
\begin{equation}
\label{eq-asymmetry-ineq}
A(\eps) \leq \amax \leq g \, .
\end{equation}
Here, $\amax$ is the unique maximum of the function $A(\eps)$. Its position and value are
\begin{align}
\eps_0 			   &= \sqrt{2 (1-g) + q^2} \, , \label{eq-eps0} \\
\amax  &= A(\eps_0) = \dfrac {\abs{q}}
			         {\sqrt{2 (1-g) + q^2}} \; g \, . \label{eq-amax}
\end{align}
Interestingly, $\amax \rightarrow g$ for $g \rightarrow 1$, such that the lower bound of $g$ by $\amax$ becomes particularly useful in the interesting regime of high coherence $g$. The bound further becomes better with increasing $\abs{q}$, which  allows for a stringent approximation also in the case of lower $g$. 
\begin{figure}[t]
	\centering
	\includegraphics[width=\columnwidth]{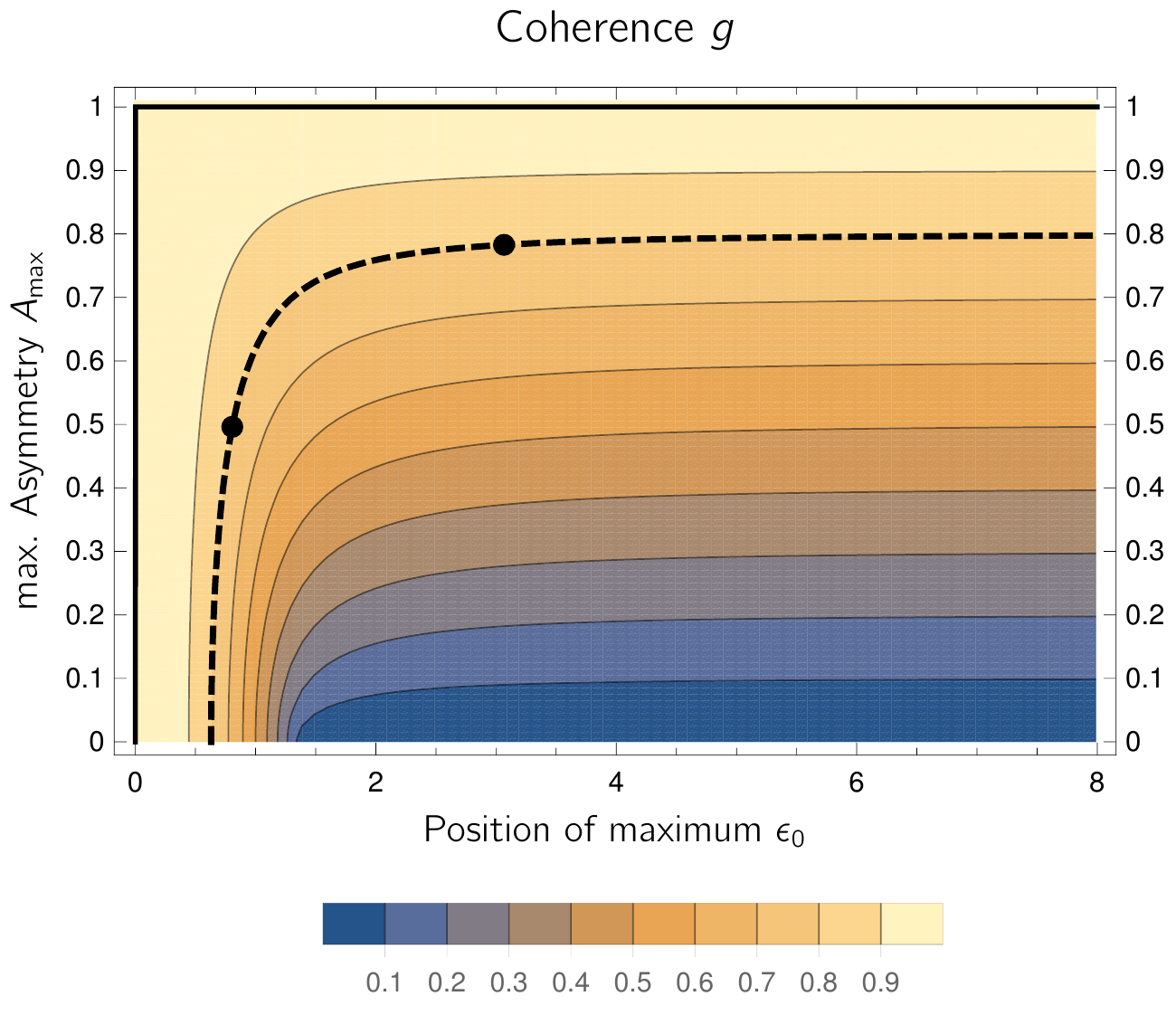}
	\caption{
	(Color online) Contour plot of $g(\eps_0, \amax)$ (Eq.~\ref{eq-g-of-amax-eps0}) against the maximum value $\amax$ and its position $\eps_0$ of the asymmetry parameter. The dashed $g= 0.8$ contour and the dots from Fig.~\ref{fig:asymmetry} are repeated for orientation. Along the contours, $|q|$ increases from left to right. Independent of $g$, the asymmetry maximum approaches $g$ for large $|q|$, towards the right-hand side of the plot.
	}
	\label{fig:isocoherence}
\end{figure}
Fig.~\ref{fig:asymmetry} shows examples of the asymmetry parameter as function of $\eps$ for $g=0.8$. The  solid lines are drawn for $q=0.5$ and $q=3$, respectively. As expected from Eq.~\eqref{eq-asymmetry-fano}, the curves' maxima form lower bounds for the coherence parameter $g=0.8$, indicated by the dotted line. The maxima of $A(\eps)$ for all other values of $|q|$ are indicated by the black dashed line. It can be seen that $\amax$ forms a lower bound to $g$ and approaches $g$ with increasing $|q|$. 

The coherence parameter $g$ can also be calculated exactly by combining the position and value of the maximum, $\eps_0$ and $\amax$. Starting from Eqs.~\eqref{eq-eps0} and \eqref{eq-amax} and eliminating $q$, we find
\begin{align}
\label{eq-g-of-amax-eps0}
&g (\eps_0, \amax) 
= \dfrac{1}{6}
\left(
	2-\eps_0^2+\dfrac{\left(\eps_0^2 -2\right)^2}{R} + R
\right) \,, \\
 &R = \; \left [54 \amax^2 \eps_0^2-\left(\eps_0^2 -2\right)^3 \right. \nonumber \\ 
 & \qquad \left. + 6 \sqrt{3}
\sqrt{27 \amax^4 \eps_0^4-\amax^2 \eps_0^2 \left(\eps_0^2 -2\right)^3}\right]^{1/3}\,.
\end{align}
This result is illustrated in Fig.~\ref{fig:isocoherence} as a contour plot, i.e. for each point $(\eps_0, \amax)$, the coherence is shown color-coded. Thus, the dashed line in Fig.~\ref{fig:asymmetry} can be found again in Fig.~\ref{fig:isocoherence} as the $g=0.8$ contour. Along the contours, $|q|$ increases from left to right. The two dots again show the maxima of the two curves of Fig.~\ref{fig:asymmetry} for $|q|= 0.5$ and $|q|= 3$, respectively. Independent of $g$, the maximum $\amax$ approaches $g$ with increasing $|q|$ on the right-hand side of Fig.~\ref{fig:isocoherence}.

{\it Discussion. }
The determination of $g$ in a FI is already possible from a single measurement of the intensity as a function of $\eps$. Over a wide range of system parameters, the approximate determination of $g$ via a lower bound Eq.~(\ref{eq-asymmetry-ineq}) may be sufficient. It does not require a normalization of the energy axis, since it only relies on the amplitude of the asymmetry parameter. The experimental uncertainty can further be reduced by combining results from spectra with different $q$, as illustrated in Fig.~\ref{fig:asymmetry}. The exact formula \eqref{eq-g-of-amax-eps0} for $g(\amax, \eps_0)$, on the other hand, works even for low $|q|$.

It is crucial to note that the standard visibility Eq.~(\ref{eq-visibility}) may strongly over- or underestimate the true coherence parameter $g$, see Fig.~\ref{fig:traditional}(e). Then, a naive application of the standard relation Eq.~(\ref{vis-g}) between visibility and coherence leads to incorrect results. This is of particular importance if, e.g., an application requires a certain minimum coherence. For example, a two-channel quantum state in a MZI violates local realism if it exhibits a sufficiently high degree of coherence and anti-correlation between the two channels~\cite{johansen_bells_1996}. 

We further note that while $A(\eps)$ is readily evaluated from a given experimental spectrum, it is important that experimental baselines are treated with care, as it is required in evaluating the standard visibility.

A baseline in the spectrum, $I(\eps) \rightarrow I(\eps) + \beta$, will reduce the asymmetry \eqref{eq-asymmetry}, so $\amax$ will still be a lower bound for $g$. However, the approximation will not be as good and the exact formula \eqref{eq-g-of-amax-eps0} will break down. The baseline of unknown origin may not be carelessly substracted, since one can show analytically that 
\begin{align}
  \alpha [\beta + \, I(\eps, q, g)] = I(\eps, q', g'=1)\,,
\end{align}
i.e., a FI spectrum with non-perfect $g<1$ can mimic a spectrum with complete coherence $g'=1$, provided that a prefactor $\alpha$ and a baseline $\beta$ are suitably chosen, and that the true $q$-factor is modified to $q'$. Here, $\alpha$, $\beta$ and $q'$ only depend on $g$ and $q$. If baseline and prefactor are unknown in an experiment, an experimental determination of $q$ can be used to mitigate this issue. Assuming that $g$ is independent of $q$, also spectra obtained for  different $q$ can be combined to improve the measurement of $g$. 

In summary, we have shown that for FI, the coherence and visibility do not satisfy the standard textbook relation known for MZI, and that the visibility cannot be used to determine the coherence. Therefore, we proposed a different approach to determine the coherence parameter, based on the asymmetry of the Fano lineshape. Like the visibility in the case of MZI, the maximum of the asymmetry parameter is a lower bound for the coherence,  and approximates it well under certain practically achievable conditions. The coherence parameter can be determined exactly from the maximal asymmetry and its position. Our results set the stage for the further development of FI towards applications involving coherence and interference phenomena, and we in particular envision the exploration of non-classical states in x-ray quantum optics.
\bibliographystyle{myprsty}
\bibliography{fanocoherence}
\end{document}